\documentclass[12pt]{article}
\usepackage{amsfonts}
\usepackage{amsmath,amssymb}
\usepackage{graphicx}

\baselineskip %
\advance %
\textheight by %
\topskip %
\makeatletter \@addtoreset{equation}{section}

\makeatother
\def\Z{\mathbb Z}
\def\N{\mathbb N}
\def\R{\mathbb R}

\def\be{\begin{equation}}
\def\ee{\end{equation}}

\def\bea{\begin{eqnarray}}

\def\obj(#1)(#2)#3#4#5{%
  \psline[arrows={[-]}, linestyle=dashed, dash=0.0 1,dashadjust=false](#1)(#2)%
  \uput{0.4}[90](#1){#3}%
  \uput{0.4}[-90](#1){#4}\uput{0.4}[-90](#2){#5}%
}
\def\objd(#1)(#2)#3#4#5{%
  \psline[arrows={[-}, linestyle=dashed, dash=0.0 1,dashadjust=false](#1)(#2)%
  \uput{0.4}[90](#1){#3}%
  \uput{0.4}[-90](#1){#4}\uput{0.4}[-90](#2){#5}%
}

\def\eea{\end{eqnarray}}

\def\ben{\begin{displaymath}}
\def\een{\end{displaymath}}
\def\ba{\begin{array}{c}}
\def\bal{\begin{array}{l}}
\def\ea{\end{array}}

\begin{document}

\title{Aharonov-Bohm effect on AdS${}_2$ and nonlinear supersymmetry of
reflectionless P\"oschl-Teller system}

\author{\textsf{Francisco Correa}\textsf{$\,$, V\'{\i}t Jakubsk\'y}
\textsf{$\,$ and  Mikhail S. Plyushchay}\\
{\small \textit{Departamento de F\'{\i}sica, Universidad de
Santiago de Chile, Casilla 307, Santiago 2, Chile}}\\
\sl{\small{E-mails:  fco.correa.s@gmail.com, v.jakubsky@gmail.com,
mplyushc@lauca.usach.cl}} }
\date{}
\maketitle

\begin{abstract}
We explain the origin and the nature of a special nonlinear
supersymmetry of a reflectionless P\"oschl-Teller system by the
Aharonov-Bohm effect for a nonrelativistic particle on the
AdS${}_2$. A key role in the supersymmetric structure appearing
after reduction by a compact generator of the AdS${}_2$ isometry is
shown to be played by the discrete symmetries related to the space
and time reflections in the ambient Minkowski space. We also observe
that a correspondence between the two quantum non-relativistic
systems is somewhat of the AdS/CFT holography nature.
\end{abstract}

\section{Introduction}

Exactly solvable quantum hyperbolic P\"oschl-Teller system
\cite{PT}--\cite{Ros3} finds diverse applications in a variety of
physical problems. It appears, for instance, in the form of a
stability equation for soliton (kink) solutions in 1+1-dimensional
$\varphi^4$ and sine-Gordon field theories \cite{Jackiw}. It
describes static solutions in the Gross-Neveu model
\cite{Dash,Fein,Dunne}, and was used in investigation of the
tachyon condensation phenomenon in gauge field string dynamics
\cite{Sen,Zwieb}. This system, on the other hand, plays a
fundamental role in the inverse scattering problem method for
nonlinear evolution equations \cite{Darb,integrable}.

All the mentioned applications are related to a peculiar property of
the P\"oschl-Teller system, whose potential $U(x)=-\lambda\cosh^{-2}
x$ becomes reflectionless at special values $\lambda=m(m+1)$,
$m=1,2,\ldots$, of the coupling constant. This property finds a
simple explanation in terms of almost isospectral transformations
underlying supersymmetric quantum mechanics \cite{susyqm}. Namely,
the reflectionless P\"oschl-Teller (RPT) system with
$\lambda=m(m+1)$ can be related to a free particle model ($m=0$) by
a Crum-Darboux transformation of order $m$. The RPT system possesses
also a shape-invariance property, due to which the energies of its
$m$ bound states are found algebraically \cite{susyqm,Levai,Rosu}.

The RPT system with $m$ bound states is characterized by an integral
of motion that is a differential operator of order $2m+1$. This
operator is known as a nontrivial operator of the Lax pair of a
non-periodic problem for the $m$-th order Korteweg-de Vries equation
\cite{integrable}. It was rediscovered recently in the context of
nonlinear supersymmetric structure associated with the RPT system,
see Refs. \cite{PTsusy,pecu}, and \cite{AIS,AndSok}. More
specifically, it was observed in \cite{PTsusy} that the RPT system
is characterized by a hidden bosonized nonlinear $N=2$ supersymmetry
\cite{Bosusy}. The nontrivial Lax operator plays the role of one of
its two supercharges, and the parity (reflection) operator $R$,
$R\psi(x)=\psi(-x)$, is identified as the $\Z_2$-grading operator
$\Gamma$. As a consequence, the extended system composed from the
two RPT systems with parameters $m$ and $l<m$ (including the case
$l=0$ that corresponds to a free particle system), is described by a
nonlinear tri-supersymmetry \cite{trisusy,trisusy1}. The
tri-supersymmetric structure  admits three alternative choices of
the grading operator, $\Gamma=\sigma_3,$ $R$, or $\sigma_3R$, where
$\sigma_3$ is a diagonal Pauli matrix \cite{tri-delta}. For
$\Gamma=\sigma_3$, the local part of this unusual supersymmetric
structure corresponds to a nonlinear $N=4$ supersymmetry, in which
the higher order integral associated with the nontrivial Lax
operator plays the role of a bosonic central charge \cite{trisusy1}.

In the present paper we show that the hidden bosonized nonlinear
supersymmetry of the RPT system and the related nonlinear
tri-supersymmetric structure of the extended case originate from a
quantum problem of a charged non-relativistic particle on the
AdS${}_2$ surface  in the presence of a singular magnetic-like
vortex. Classically, the particle performs a free, geodesic motion
on the AdS${}_2$.  Aharonov-Bohm effect \cite{AhB}
 influences essentially on the  quantum properties  of the
system. For half-integer values of the flux, the spectrum is
characterized by a specific additional double degeneration related
to an involutive automorphism of  the AdS${}_2$ isometry.  The
system reveals a reflectionless  quantum dynamics in this case.
These two related properties underlie the peculiar nonlinear
supersymmetric structure of the RPT system obtainable by an
appropriate angular momentum reduction. At the same time for a
generic magnetic flux value, the Aharonov-Bohm effect on the
AdS${}_2$ provides a new vision on the nature of the Crum-Darboux
transformations for the quantum hyperbolic P\"oschl-Teller system in
the context of the AdS/CFT correspondence \cite{ads1,ads4}. We also
discuss this aspect in the background of non-relativistic AdS/CFT.
The latter is based on non-relativistic conformal symmetry
\cite{NiedHag}--\cite{LP1}, and became recently a hot topic due to
the diverse range of applications, see \cite{NRCFT,DHH}.

The paper is organized as follows. In Section 2 we summarize briefly
the results on the nonlinear hidden bosonized supersymmetric
structure of the RPT system and the tri-supersymmetric structure of
the pair of such systems in the light of the Crum-Darboux
transformations, discussed in detail in \cite{PTsusy,trisusy1}. In
Section 3 we investigate the classical and quantum theory of a
non-relativistic particle on the AdS${}_2$, minimally coupled to a
U(1) gauge field given by the Aharonov-Bohm vector potential. In
Section 4 we show how the nonlinear supersymmetric structure of the
RPT system emerges from such a non-relativistic system at
half-integer values of the magnetic flux. In Section 5 we discuss a
realization of the conformal and super-conformal dynamical
symmetries in the RPT system. We conclude with a discussion of the
results and open research questions in Section 6.

\section{Crum-Darboux transformations and tri-supersym-\\metry}

A hyperbolic reflectionless  P\"oschl-Teller system is described by
the Hamiltonian\
\begin{equation}
    H_m=-\frac{d^2}{dx^2}- \frac{m(m+1)}{\cosh^2 x}\, .
    \label{PT}
\end{equation}
Here and in what follows we put $\hbar=1$. It is convenient to treat
$m$ as a parameter that can take any integer value, $m\in \Z$. Then
$m=0$ corresponds to a free particle case, and we have the identity
\begin{equation}\label{Hmm}
    H_m=H_{-(m+1)}.
\end{equation}
In terms of the first order differential operators
\begin{equation}
    \mathcal{D}_{m}=\frac{d}{d x}+m\tanh x, \qquad
     \mathcal{D}_{-m}=-\mathcal{D}^{\dagger}_{m}\,,\label{D}
\end{equation}
the second order operator (\ref{PT}) can be presented in a form
\begin{equation}\label{HmmD}
    H_m=-\mathcal{D}_{-m}\mathcal{D}_{m}-m^2,
\end{equation}
while (\ref{Hmm}) rewrites equivalently
\begin{equation}\label{shape}
    \mathcal{D}_{-m}\mathcal{D}_{m}=\mathcal{D}_{m+1
    }\mathcal{D}_{-m-1}+(2m+1).
\end{equation}
Using this identity  and (\ref{HmmD}), one can check that there hold
the following intertwining relations
\begin{equation}\label{Darb}
     \mathcal{D}_{m}H_{m}=H_{m-1}\mathcal{D}_{m}, \qquad
     \mathcal{D}_{-m}H_{m-1}=H_{m}\mathcal{D}_{-m}.
\end{equation}
These relations can be understood from a point of view of 
Crum-Darboux theorem~\cite{Darb}. According to it, if a
differential operator $A_n$  of order $n$ annihilates $n$
eigenstates $\psi_i$ of a Hamiltonian $H$, one can construct
another Hamiltonian $\tilde{H}=H-2(\ln
W(\psi_1,\ldots,\psi_n))''$, and these three operators are related
by the identities
\begin{equation}\label{tilH}
    \tilde{H}A_n=A_nH,\qquad
    A_n^\dagger\tilde{H}=HA_n^\dagger.
\end{equation}
Here, $W$ is the Wronskian of not obligatorily to be physical states
$\psi_i$, $i=1,\ldots, n$, such that $W\neq 0$. As a consequence of
identities (\ref{tilH}), if $\psi$ is an eigenstate of
$H$, $H\psi=E\psi$, $A_n\psi\neq 0$, the state $A_n\psi$ will be the
eigenstate of $\tilde{H}$  with the same eigenvalue $E$.  And vice
versa, if $\tilde{\psi}$ is an eigenstate of $\tilde{H}$,
$\tilde{H}\tilde{\psi}=\tilde{E}\tilde{\psi}$,  such that
$A_n^\dagger \tilde{\psi}\neq 0$, the state
$A_n^\dagger\tilde{\psi}$ will be an eigenstate of $H$ with the the
same eigenvalue $\tilde{E}$. This means that the Hamiltonians $H$
and $\tilde{H}$ will be almost isospectral. In the case of (\ref{Darb}), the first
order operator ${\cal D}_m$ annihilates the nodeless ground state
$\cosh^{-m}x$ of the Hamiltonian $H_m$ with eigenvalue $-m^2$, and
$H_{m-1}$ has the same spectrum as $H_m$ except the missing in it
eigenvalue $-m^2$. In addition, relations (\ref{Darb}) reflect the
shape-invariance property of the P\"oschl-Teller system. The first
order Darboux transformation generated by the operator ${\cal D}_m$
produces for the Hamiltonian $H=H_m$ the partner Hamiltonian
$\tilde{H}$, which has potential of the same form but with the
parameter $m$ shifted in one, i.e. $\tilde{H}=H_{m-1}$.

By repeated application of (\ref{Darb}), we can get the intertwining
relations of higher order Crum-Darboux transformations \cite{Darb},
\begin{eqnarray}
     &\big(\mathcal{D}_{m-l}\ldots \mathcal{D}_{m-1}\mathcal{D}_{m}\big)H_{m}=H_{m-l-1}
     \big(\mathcal{D}_{m-l}\ldots
     \mathcal{D}_{m-1}\mathcal{D}_{m}\big),&\label{CD1}\\
     &\big(\mathcal{D}_{-m}\mathcal{D}_{-(m-1)}
     \ldots \mathcal{D}_{-(m-l)}\big)H_{m-l-1}=
     H_{m}\big(\mathcal{D}_{-m}\mathcal{D}_{-(m-1)}\ldots
     \mathcal{D}_{-(m-l)}\big).&\label{CD}
\end{eqnarray}
In a particular case $l=m-1$, relation (\ref{CD}) takes a form
\begin{equation}\label{H0}
    \big(\mathcal{D}_{-m}\mathcal{D}_{-(m-1)}
     \ldots \mathcal{D}_{-1}\big)H_{0}=
     H_{m}\big(\mathcal{D}_{-m}\mathcal{D}_{-(m-1)}\ldots
     \mathcal{D}_{-1}\big),
\end{equation}
which means that the system (\ref{PT}) is almost isospectral to
the free particle system. Making use of it, the scattering states
of the RPT system (\ref{PT}) can be obtained from the plane wave
eigenstates of the free particle,
\begin{equation}\label{PTcont}
    \psi _{m}^{(\pm k)}(x)=\mathcal{D}_{-m}
    {\cal D}_{-m+1}\ldots \mathcal{D}_{-1}\cdot
    \exp (\pm ik x).
\end{equation}
According to (\ref{H0}) and (\ref{PTcont}), positive energy values
of the system (\ref{PT}), $E_{m,k}=k^2$ with $k> 0$, are doubly
degenerate, while zero energy with $k=0$ corresponds to a singlet
state
\begin{equation}\label{E=0}
    \psi _{m}^{(\pm 0)}( x)\equiv \psi_{m;m}(x)=\big(\mathcal{D}_{-m}
    {\cal D}_{-m+1}\ldots \mathcal{D}_{-1}\big)\cdot 1.
\end{equation}

Function $\cosh^{-m}x$ is the zero mode of the first order operator
$\mathcal{D}_{m}$. In correspondence with (\ref{HmmD}), it describes
a bound state of the RPT system with energy $E_{m;0}=-m^2$.  This
observation  together with the relation (\ref{CD}) taken for
$l=0,\ldots,m-2$ allow us to find the whole set of the of $m$ bound
singlet states of the RPT system (\ref{PT}),
\begin{equation}\label{nml}
    \psi_{m;0}(x)=\cosh^{-m}x, \quad \psi_{m;n}(x)={\cal
    D}_{-m}{\cal D}_{-m+1}\ldots
    {\cal D}_{-m+n-1}\cosh^{n-m}x, \quad n=1,\ldots, m-1.
\end{equation}
Extension of (\ref{nml}) for $n=m$ ($l=m-1$) reproduces singlet
state (\ref{E=0}) of the continuous part of the spectrum. The
energies of all the $m+1$ singlet states (\ref{nml}) and
(\ref{E=0}) are given then by
\begin{equation}
    E_{m;n}=-(m-n)^2,\qquad n=0,\ldots, m.
    \label{singlet}
\end{equation}

The double degeneration in the continuous part of the spectrum and
the presence of  $m+1>1$ singlet states indicate on a hidden,
nonlinear supersymmetry  in the system (\ref{PT}). Its
corresponding supercharges can be identified easily. Applying the
order $2m+1$ Crum-Darboux transformation (\ref{CD}) corresponding
to $l=2m$ and taking into account  relation (\ref{Hmm}), we find
that the RPT system (\ref{PT}) is characterized by a local
integral
\begin{equation}\label{calA}
    \mathcal{A}_{2m+1}=\mathcal{D}_{-m}
    \mathcal{D}_{-m+1}\ldots \mathcal{D}_{0}\ldots
    \mathcal{D}_{m-1}\mathcal{D}_{m},\qquad
    [\mathcal{A}_{2m+1},H_{m}]=0,
\end{equation}
that is a differential operator of order $2m+1$. This is a
nontrivial integral of the Lax pair $({\cal A}_{2m+1},H_m)$ of the
$m$-th order KdV equation \cite{integrable}. It is a parity-odd
operator, while the Hamiltonian (\ref{PT}) is parity-even.
Identifying reflection (parity) operator $R$, $R\psi(x)=\psi(-x)$,
as a grading operator, and integrals
$Z_1=i^{2m+1}\mathcal{A}_{2m+1}$ and $Z_2=iRZ_1$ as Hermitian
supercharges, we find that the RPT system (\ref{PT}) is
characterized by a nonlinear $N=2$ supersymmetry
\begin{equation}
    [Z_a,H_m]=0, \qquad
    \{Z_a,Z_b\}=2\delta_{ab}P_{2m+1}(H_m), \label{bososusy}
\end{equation}
where $P_{2m+1}(H_m)$ is a polynomial of order $2m+1$. Its explicit
form can be found with the help of relation (\ref{shape}),
\begin{equation}\label{P2m+a}
    P_{2m+1}(H_m)=(H_m-E_{m;m})\prod_{n=0}^{m-1}(H_m-E_{m;n})^2,
\end{equation}
where $E_{m;n}$ are the energies (\ref{singlet}) of the singlet
states. Singlet states (\ref{E=0}) and (\ref{nml}) are zero modes
of the supercharges $Z_a$,  other $m$ states annihilated by $Z_a$
have a more intricate nature, see
 \cite{pecu,trisusy}.

Note that the nontrivial integral (\ref{calA}) has a sense of the
integral ${\cal D}_0$ of the free particle system $H_0$
transferred to the RPT system $H_m$  by means of the Crum-Darboux
transformations. Indeed, multiply the relation $H_0{\cal
D}_0={\cal D}_0H_0$ from the left by the operator ${\cal
D}_{-m}{\cal D}_{-m+1} \ldots {\cal D}_{-1}$,  and from the right
by ${\cal D}_1\ldots {\cal D}_{m-1}{\cal D}_m$. Using on the left
hand side intertwining relation (\ref{H0}), and on the right hand
side its Hermitian conjugate form, we get $[{\cal
A}_{2m+1},H_m]=0$.

The system composed from the two RPT systems $H_m$ and $H_l$, $0\leq
l<m$,  can be described by the $2\times 2$ matrix Hamiltonian
\begin{equation}
    \mathcal{H}_{m,l}=\left(
    \begin{array}{cc}
    {H}_{l} & 0 \\
    0 & {H}_{m}
    \end{array}
    \right).
    \label{superPT}
\end{equation}
The two RPT subsystems are almost isospectral. The subsystem $H_l$
does not have $m-l$ energy levels corresponding to the $m-l$
lowest bound states of the subsystem $H_m$. Hence, these states
form singlets of the extended system.
 The intertwining relations (\ref{CD}) allow us to identify
the local Hermitian integrals of motion of the extended system
(\ref{superPT}),
 \begin{equation}
     \mathcal{X}_{m,l}=
    \left(\begin{array}{cc} 0&X^-_{m,l}\\X^+_{m,l}&0
    \end{array}\right),\quad
    \mathcal{Y}_{m,l}=
    \left(\begin{array}{cc} 0&Y^-_{m,l}\\Y^+_{m,l}&0\end{array}\right),\quad
    \mathcal{Z}_{m,l}=
    \left(\begin{array}{cc} Z^+_{m,l}&0\\0&Z^-_{m,l}
    \end{array}\right),
    \label{maY}
\end{equation}
$[\mathcal{H}_{m,l},\mathcal{X}_{m,l}]=[\mathcal{H}_{m,l},
    \mathcal{Y}_{m,l}]=[\mathcal{H}_{m,l},\mathcal{Z}_{m,l}]=0,
$ where
\begin{eqnarray}
  X^-_{m,l}&=&(X^+_{m,l})^\dagger =-i^{m-l}
  \mathcal{D}_{l+1}\mathcal{D}_{l+2}\ldots\mathcal{D}_{m-1}\mathcal{D}
    _{m},\label{DX}\\
  Y^-_{m,l}&=&(Y^+_{m,l})^\dagger=i^{2l+1}\mathcal{A}_{2l+1}X^-_{m,l},\label{DY}\\
     Z^-_{m,l}&=&(Z^-_{m,l})^\dagger=X^+_{m,l}Y^-_{m,l}=Y^+_{m,l}X^-_{m,l}=
    i^{2m+1}\mathcal{A}_{2m+1},\label{Zm}\\
    Z^+_{m,l}&=&(Z^+_{m,l})^\dagger=X^-_{m,l}Y^+_{m,l}=Y^-_{m,l}X^+_{m,l}=
    i^{2l+1}X^-_{m,l}X^{+}_{m,l}\mathcal{A}_{2l+1}\, .\label{Z+ml}
\end{eqnarray}
These integrals  of motion mutually commute,
\begin{equation}\label{XYZ}
    [\mathcal{X}_{m,l},\mathcal{Y}_{m,l}]=[\mathcal{Y}_{m,l},\mathcal{Z}_{m,l}]=
    [\mathcal{X}_{m,l},\mathcal{Z}_{m,l}]=0,
\end{equation}
and satisfy relations
\begin{equation}\label{ZXY}
    \mathcal{Z}_{m,l}=\mathcal{X}_{m,l}\mathcal{Y}_{m,l},
\end{equation}
\begin{equation}
   (\mathcal{X}_{m,l})^2=
   P_\mathcal{X}(\mathcal{H}_{m,l}), \qquad
     (\mathcal{Y}_{m,l})^2=P_\mathcal{Y}(\mathcal{H}_{m,l}),
     \qquad
     (\mathcal{Z}_{m,l})^2=P_\mathcal{Z}(\mathcal{H}_{m,l}).
   \label{XXYY}
\end{equation}
In correspondence with (\ref{ZXY}), $
    P_\mathcal{Z}(\mathcal{H}_{m,l})=P_\mathcal{X}(\mathcal{H}_{m,l})
     P_\mathcal{Y}(\mathcal{H}_{m,l})
$
is a polynomial of order $2m+1$ of the form (\ref{P2m+a}) with $H_m$
changed for  $\mathcal{H}_{m,l}$, and
\begin{equation}\label{PXH}
    P_\mathcal{X}(\mathcal{H}_{m,l})=\prod_{n=0}^{m-l-1}(\mathcal{H}_{m,l}-
    E_{m;n}),\quad
    P_\mathcal{Y}(\mathcal{H}_{m,l})=P_\mathcal{X}(\mathcal{H}_{m,l})\cdot
    (\mathcal{H}_{m,l}-E_{m;m})\prod_{n=m-l}^{m-1}(\mathcal{H}_{m,l}-
    E_{m;n})^2.
\end{equation}
Here, the roots $E_{m;n}$ of the polynomials correspond to the
singlet states of the subsystem $H_m$, and are given by
(\ref{singlet}).

 In addition to integrals (\ref{maY}), the extended system
(\ref{superPT}) has also three mutually commuting \emph{trivial}
integrals of motion,
\begin{equation}
\Gamma_i \in \{ R,\sigma_3,R\sigma_3 \}, \qquad \Gamma_i^2=1,\quad
i=1,2,3. \label{trivial}
\end{equation}
Any of them can be chosen as a grading operator $\Gamma$, that
classifies one of the integrals (\ref{maY}) as an even (bosonic),
and the other two as odd (fermionic) generators. As a result, the
set of integrals (\ref{maY}), (\ref{trivial}) extended by the
products of (\ref{trivial}) with (\ref{maY}) generates together
with Hamiltonian (\ref{superPT}) a certain deformation of the
$su(2|2)$ superalgebra. Its concrete form depends on the choice of
the grading operator, see \cite{trisusy,trisusy1}. For the
particular choice $\Gamma=\sigma_3$, the subset of integrals
$\mathcal{X}_{m,l}$, $\mathcal{Y}_{m,l}$,
$i\sigma_3\mathcal{X}_{m,l}$, and $i\sigma_3\mathcal{Y}_{m,l}$ is
identified as the set of the odd supercharges, while
$\mathcal{Z}_{m,l}$ and $\mathcal{H}_{m,l}$ are even operators.
These \emph{local} integrals generate a nonlinear $N=4$
superalgebra, in which, in correspondence with (\ref{XYZ}),
$\mathcal{Z}_{m,l}$ plays a role of a central charge. For $m-l=1$,
the integral $\mathcal{X}_{m,m-1}$ is a usual first order
supercharge for the superextended system (\ref{superPT}), whose
conservation follows from the Darboux intertwining relations
(\ref{Darb}). In a generic case, the relation of commutativity of
$\mathcal{X}_{m,l}$ with matrix Hamiltonian $ \mathcal{H}_{m,l}$
represents just mutually conjugate  intertwining relations (\ref{CD1}) and
(\ref{CD}) of order $m-l$. The integral $\mathcal{Y}_{m,l}$
corresponds to the Crum-Darboux transformations of order
$m+l+1$ that intertwine the systems $H_m$ and $H_{-(l+1)}=H_l$.

\section{Aharonov-Bohm effect on the AdS${}_2$}

In this section, we will consider a two-dimensional setting with the
geometry of the AdS${}_2$ space. The particle system will be studied
in the presence of a singular magnetic flux in both classical and
quantum frameworks. This will provide a valuable background for
deeper understanding of the algebraic properties described in the
previous section.

Consider a one-sheeted hyperboloid
\begin{equation}\label{hyp}
    x^\mu x_\mu = -x_1^2-x_2^2+x_3^2=-{\cal R}^2
\end{equation}
embedded in a three-dimensional Minkowski space with metric
$ds^2=dx^\mu dx^\nu\eta_{\mu\nu}$, $\eta_{\mu\nu}=diag\,
(-1,-1,+1)$, $\mu,\nu=1,2,3$. It  can be parameterized by
\begin{equation}
    x^1 =  {\cal R} \cosh \chi \cos \varphi,\quad
    x^2 = {\cal R} \cosh \chi  \sin\varphi,\quad
    x^3 = {\cal R} \sinh \chi,
    \label{coord}
\end{equation}
and is identified as the AdS${}_2$ space of radius ${\cal R}>0$
with the induced metric $ds^2={\cal R}^2(d\chi^2-\cosh^2\chi
d\varphi^2)$, $-\infty<\chi<\infty$, $0\leq \varphi<2\pi$
\footnote{The AdS${}_2$ has a topology $\R\times S^1$ with $S^1$
usually unwrapped \cite{ads4}. However, for here, the
$S^1$-topology plays a key role.}. Let in an ambient Minkowski
space the Aharonov-Bohm vector potential is given,
\begin{equation}
    A_1=-\frac{\Phi }{2\pi}\frac{x_2}{x_1^2+x_2^2}, \quad A_2=\frac{\Phi
    }{2\pi}\frac{x_1}{x_1^2+x_2^2}, \quad A_3=0,\label{AB}
\end{equation}
that describes a ``magnetic" field of a flux line along the
$x_3$-axis, $B_\mu=\epsilon_{\mu\nu\lambda}\partial^\nu
A^\lambda=(0,0,\Phi\delta^2(x_1,x_2))$. Here
$\epsilon_{\mu\nu\lambda}$ is an antisymmetric tensor,
$\epsilon_{123}=1$.

\subsection{Classical system}

Consider  a \emph{non-relativistic} charged particle minimally
coupled to the external U(1) gauge field (\ref{AB}), and confined to
move on the two-dimensional surface (\ref{hyp}). Taking into
account Eqs. (\ref{coord}), (\ref{AB}), the corresponding Lagrangian
of a particle of unit mass,
$L=\frac{1}{2}\dot{x}_{\mu}\dot{x}^{\mu}+\frac{e}{c}A_{\mu}\dot{x}^{\mu}$,
is reduced to
\begin{equation}\label{lanchi}
    L=
    \frac{{\cal R}^2}{2}(\dot{\chi}^2-\cosh^2\chi\,
    \dot{\varphi}^2)-\alpha\dot{\varphi},
\end{equation}
where $\alpha\equiv \frac{e\Phi}{2\pi c}$,
$\dot{x}{}^\mu=dx^\mu/dt$, and $t$ is an evolution parameter. The
last,  coupling term in (\ref{lanchi}) is a total time derivative.
It does not effect on the classical dynamics of the particle that
performs a geodesic motion on the AdS${}_2$.

The form of the trajectories can be identified by noting that the
SO(2,1)-isometry of the AdS${}_2$ is a transitive symmetry group
of the hyperboloid (\ref{hyp}), i.e. any two points on the surface
can be related by an appropriate SO(2,1)-transformation. The
generators of this symmetry  are the integrals of motion
\begin{eqnarray}
    &J_1= -p_\chi \sin \varphi -J_3\cos \varphi \tanh \chi,\quad
    J_2=p_\chi \cos \varphi - J_3\sin \varphi \tanh
    \chi,&\label{JJ12}\\
    &J_3=p_{\varphi}+\alpha,& \label{J3al}
\end{eqnarray}
where $
    p_{\chi}=\frac{\partial L}{\partial \dot{\chi}}$,
    $p_{\varphi}=\frac{\partial L}{\partial
    \dot{\varphi}}$
are the canonical momenta, $\{\varphi,p_{\varphi}\}=1$,
$\{\chi,p_\chi\}=1$. With respect to the Poisson brackets, the
integrals (\ref{JJ12}) and (\ref{J3al}) generate the (2+1)D Lorentz
algebra
\begin{equation}\label{so21clas}
    \{J_\mu,J_\nu\}=-\epsilon_{\mu\nu\lambda}J^\lambda.
\end{equation}
Their conservation follows from the form of the canonical
Hamiltonian, $H=p_{\chi}\dot{\chi}+p_{\varphi}\dot{\varphi}-L$,
\begin{equation}
    H=\frac{1}{2{\cal R}^2}\left(p^2_{\chi}-
    \frac{(p_{\varphi}+\alpha)^2}{\cosh^{2}\chi}\right),
    \label{H}
\end{equation}
that is reduced to the $so(2, 1)$  Casimir element up to a multiplicative constant
\begin{equation}\label{HJ}
    H=-\frac{1}{2{\cal R}^2}{\cal C},\quad {\cal C}=J_{\mu}J^{\mu}.
\end{equation}
Direct checking shows that the integrals (\ref{JJ12}),
(\ref{J3al}) satisfy identically the relation $x^\mu J_\mu =0$
with $x^\mu$ given by (\ref{coord}). This also follows from the
observation that the system (\ref{H}) can alternatively be
obtained in two steps. First,  we reduce the $ISO(2,1)$-invariant
free particle system $L=\frac{1}{2}m\dot{x}_\mu \dot{x}^\mu$ to
the $SO(2,1)$-invariant surface given by the second class
constraints $x_\mu x^\mu+{\cal R}^2=0$ and $p_\mu x^\mu=0$ in the
phase space with canonical coordinates ($x^\mu$, $p_\mu$). The
reduced phase space is described by the two pairs of canonical
variables $(\varphi$, $p_\varphi$), ($\chi$, $p_\chi$), in terms
of which the $SO(2,1)$ generators
$J_\mu=-\epsilon_{\mu\nu\lambda}x^\nu p^\lambda$ take the form of
the integrals (\ref{JJ12}), (\ref{J3al}) with $\alpha=0$. The case
$\alpha\neq 0$ is obtained then via a subsequent canonical
transformation $p_{\varphi}\rightarrow p_{\varphi}+\alpha$. Both
steps do not touch the initial free case identity $x^\mu
J_\mu\equiv 0$. As a result, the trajectory is determined by the
intersection of the Minkowski hyperplane $x^{\mu}J_{\mu}=0$,
$J_\mu=const$,  with the surface of the hyperboloid (\ref{hyp}).
Its form depends on the value of the Casimir ${\cal C}$. For
${\cal C}>0$, $=0$, or $<0$, the trajectory is respectively an
ellipse, a straight line, or a hyperbola.

Due to the Lorentzian (indefinite) metric of the AdS${}_2$ surface,
the values of the Hamiltonian (\ref{H}) are not restricted from
below. We will be interested in  the quantum system reduced to
certain levels of the integral $J_3$, in which the spectrum is
bounded from below. As it follows from  (\ref{J3al}) and (\ref{H}),
such a reduced system  describes a one-dimensional P\"oschl-Teller
system. Three types of the classical trajectories of the system
(\ref{H}) correspond to a bounded periodic (${\cal C}>0$), or unbounded
(${\cal C}\leq 0$) particle motions in a 1D \emph{classical
attractive} potential. In particular, straight line trajectories on
the hyperboloid correspond to a zero energy motion of the
P\"oschl-Teller system.

Due to its algebraic background, the system (\ref{HJ}) has also two
discrete symmetries. They have no significant role in the
classical theory, but will be of a key importance at the quantum
level in the context of the supersymmetry we discuss. These
symmetries are the involutive automorphisms of the $so(2,1)$ algebra
(\ref{so21clas}),
\begin{equation}\label{autoSO}
    R:\, (J_1,J_2,J_3)\rightarrow (-J_1,-J_2,J_3),\qquad
    S:\, (J_1,J_2,J_3)\rightarrow (-J_1,J_2,-J_3).
\end{equation}
In the ambient Minkowski space they correspond to a change of a
sign of a space-like coordinate $x^3$, and of one of the time-like
coordinates which, in correspondence with the chosen definition of
$S$, we identify with $x^2$,
\begin{equation}\label{RSxyz}
    R:\, (x^1,x^2,x^3)\rightarrow (x^1,x^2,-x^3),\qquad
    S:\, (x^1,x^2,x^3)\rightarrow (x^1,-x^2,x^3).
\end{equation}
In the curvilinear coordinates this corresponds to
\begin{equation}\label{Rchiphi}
    R:\, (\chi,\varphi)\rightarrow (-\chi,\varphi),\qquad
    S:\, (\chi,\varphi)\rightarrow (\chi,-\varphi).
\end{equation}
Lagrangian (\ref{lanchi}) is invariant under the  discrete
symmetry $R$.  However, it is quasi-invariant under symmetry $S$,
which provokes a change for a total time derivative term, $\Delta
L=2\alpha \dot{\varphi}$, that does not effect on the classical
motion. In correspondence with such a quasi-invariance of
Lagrangian, in order to reproduce involutive automorphism $S$
defined by  (\ref{autoSO}) in the presence of nonzero
Aharonov-Bohm flux $\alpha$, a transformation $\varphi\rightarrow
-\varphi$ has to be accompanied by an additional canonical
transformation $p_\varphi\rightarrow p_\varphi -2\alpha$. At the
quantum level corresponding unitary transformation is generated by
the operator $U_\alpha(\varphi)=\exp (-2i\alpha\varphi)$. This is
a well defined, $2\pi$-periodic operator only in the case of
integer and half-integer values of the flux. For $2\alpha\notin
\Z$,  as we shall see, discrete symmetry $S$ is spontaneously
broken.

\subsection{Quantization and spectral properties}

A canonical quantization of the system with a prescription of a
\emph{symmetric ordering} of non-commuting factors in quantum analog
of (\ref{JJ12}) results in the operators
\begin{eqnarray}
    \hat{J}_1&=&i\sin
    \varphi \left(\partial_\chi-\frac{1}{2}\tanh \chi \right)-\cos
    \varphi \tanh \chi \hat{J}_3,\label{Jsym1}\\
    \hat{J}_2&=&-i\cos \varphi \left(\partial_\chi-\frac{1}{2}\tanh
    \chi \right)-\sin \varphi \tanh \chi \hat{J}_3,\label{Jsym2}
    \\
    \hat{J}_3&=&-i\partial_\varphi+\alpha.\label{Jsym3}
\end{eqnarray}
They generate the $so(2,1)$ algebra,
\begin{equation}\label{so21}
    [\hat{J}_\mu,\hat{J}_\nu]=-i\epsilon_{\mu\nu\lambda}\hat{J}{}^\lambda.
\end{equation}
We assume that these operators  act on the space of the
$2\pi$-periodic in $\varphi$ wave functions $\psi(\chi,\varphi)$, where they are Hermitian with respect to a scalar product
\begin{equation}\label{scalarp}
    (\psi_1,\psi_2)=\frac{1}{2\pi}\int_{-\infty}^\infty\int_0^{2\pi}
\psi_1^*(\chi,\varphi)\psi_2(\chi,\varphi)d\chi d\varphi.
\end{equation}
The quantum Hamiltonian
\begin{equation}
    \hat{H}=-\partial_{\chi}^2-\frac
    {\hat{J}_3^2-\frac{1}{4}}{\cosh^2\chi}
    \label{Hquant}
\end{equation}
is obtained then from (\ref{HJ}), where we put $2{\cal R}^2=1$, and
subtract a  quantum constant term $\frac{\hbar^2}{4}$, i.e. we take
\begin{equation}\label{Hh}
    \hat{H}=-\hat{J}_\mu\hat{J}^\mu-\frac{1}{4}.
\end{equation}

Note that the same realization of the $so(2,1)$ generators and of
the Hamiltonian are obtained if to proceed from the definition $
\hat{J}_\mu=-\epsilon_{\mu\nu\lambda}x^\nu
\left(-i\partial^\lambda - \frac{e}{c}A^\lambda(x)\right)$ written
in the pseudospherical coordinates $({\cal R},\chi,\varphi)$,
${\cal R}>0$. In this way we get the $so(2,1)$ generators in the
form
\begin{equation}\label{J**}
    \tilde{J}_1=i\sin\varphi
    \partial_{\chi}-\cos\varphi\tanh\chi
    \hat{J}_3,\qquad
    \tilde{J}_2=-i\cos\varphi
    \partial_{\chi}-\sin\varphi\tanh\chi \hat{J}_3,\qquad
    \tilde{J}_3=\hat{J}_3.
\end{equation}
Instead of (\ref{Hquant}), with the same quantum constant shift,
we obtain
\begin{equation}\label{H*}
    \tilde{H}=-\partial_{\chi}^2-\tanh\chi\partial_{\chi}-
    (\tilde{J}_3^2-\frac{1}{4})\cosh^{-2}\chi.
\end{equation}
Operators
$\tilde{J}_\mu$ and $\tilde{H}$, $[\tilde{J}_\mu,\tilde{H}]=0$,
are Hermitian with respect to a scalar product
\begin{equation}\label{sp*}
    (\tilde{\psi}_1,\tilde{\psi}_2)=(2\pi)^{-1}\int_{-\infty}^\infty \int_0^{2\pi} \tilde{\psi}_1^*(\chi,\varphi)
    \tilde{\psi}_2(\chi,\varphi)\cosh \chi d\chi d\varphi.
\end{equation}
A subsequent similarity transformation $\tilde{\psi}\rightarrow
f\tilde{\psi}=\psi$, $\tilde{J}_\mu\rightarrow f\tilde{J}_\mu
f^{-1}=\hat{J}_\mu$, $\tilde{H}\rightarrow
f\tilde{H}f^{-1}=\hat{H}$ with $f=\sqrt{\cosh \chi}$ reduces the
scalar product (\ref{sp*}), the $so(2,1)$ generators (\ref{J**}),
and the Hamiltonian (\ref{H*}) to (\ref{scalarp}),
(\ref{Jsym1})--(\ref{Jsym3}), and (\ref{Hquant}), respectively.

Since Hamiltonian (\ref{Hquant}) is the (shifted) $so(2,1)$
Casimir operator, one can choose a representation in which
$\hat{H}$ and the compact $so(2,1)$ generator $\hat{J}_3$ are
diagonal. The stationary  Schr\"odinger equation associated with
the Hamiltonian (\ref{Hquant}) is separable in the variables
$\chi$ and $\varphi$. The common wave functions of $\hat{H}$ and
$\hat{J}_3$ can be factorized as
\begin{equation}\label{psiEm}
    \Psi^{\alpha}_{E,m}(\chi,\varphi)=e^{im\varphi}\psi^\alpha_{E,m}(\chi),\qquad
    m=0,\pm 1,\pm 2,\ldots,
\end{equation}
where the superscript marks the value of the Aharonov-Bohm flux.
 Then we have
\begin{equation}\label{j3al}
    \hat{J}_3\Psi^\alpha_{E,m}(\chi,\varphi)=j_3\Psi^\alpha_{E,m}(\chi,\varphi),\qquad j_3=m+\alpha,
\end{equation}
and the Schr\"odinger equation
\begin{equation}
    \hat{H}\Psi^\alpha_{E,m}(\chi,\varphi)=E\Psi^\alpha_{E,m}(\chi,\varphi)
\label{HPTem}
\end{equation}
is reduced to
\begin{equation}
    H_{m_\alpha}\psi^\alpha_{E,m}(\chi)=E\psi^\alpha_{E,m}(\chi),\qquad
    H_{m_\alpha}=-\frac{d^2}{d\chi^2}- \frac{m_\alpha(m_\alpha+1)}{\cosh^2
    \chi},\qquad
    m_\alpha\equiv m+\alpha-\frac{1}{2}\,.
    \label{Htmal}
\end{equation}
The reduced Hamiltonian $H_{m_\alpha}$ is just the P\"oschl-Teller
Hamiltonian (\ref{PT}) with parameter $m$ and variable $x$ changed
for $m_\alpha$ and $\chi$. Since the cases corresponding to the
Aharonov-Bohm fluxes $\alpha_1$ and  $\alpha_2=\alpha_1+n$, $n\in
\Z$, are related by a unitary transformation generated by the
operator $U_{\alpha_2,\alpha_1}(\varphi)=e^{in\varphi}$,  in what
follows we assume without loss of generality that $0\leq
\alpha<1$.

Define the linear combinations of the generators
$\hat{J}_+=\hat{J}_1+i\hat{J}_2$ and $\hat{J}_-=\hat{J}_+^\dagger$,
\begin{equation}
    \hat{J}_+=e^{i\varphi}\left(\frac{\partial}{\partial
    \chi}-\left(\hat{J}_3
    +\frac{1}{2}\right)\tanh\chi\right),\,\,
    \hat{J}_-=e^{-i\varphi}\left(-\frac{\partial}{\partial
    \chi}-\left(\hat{J}_3
    -\frac{1}{2}\right)\tanh\chi\right).\label{J+-}
\end{equation}
These are the ladder operators of the $so(2,1)$ symmetry algebra,
$[\hat{J}_3,\hat{J}_\pm]=\pm \hat{J}_\pm$,
$[\hat{J}_+,\hat{J}_-]=-2\hat{J}_3$. Then for eigenstates
(\ref{psiEm}) we have a relation
\begin{equation}\label{ladder}
    \hat{J}_3(\hat{J}_\pm\Psi^\alpha_{E,m})=(j_3\pm
    1)(\hat{J}_\pm\Psi^\alpha_{E,m}).
\end{equation}
On a subspace with $j_3=m+\alpha$, the $\chi$-dependent parts of
the ladder operators $\hat{J}_-$ and $\hat{J}_+$  correspond to
the intertwining operators (2.3). The values of the parameter $m_\alpha$
are non-integer in the case $\alpha\neq \frac{1}{2}$.

Let us discuss the spectral properties of the system in dependence
on the strength of the magnetic flux. First, consider the case
$\alpha\neq \frac{1}{2}.$ {}From the form of the reduced
Hamiltonian (\ref{Htmal}) it follows that the quantum system
(\ref{Hquant}) contains the P\"oschl-Teller subsystem with
repulsive potential $U(\chi)=+\gamma^2 \cosh^{-2}\chi$,
$\gamma^2>0$. For $\alpha=0$ this happens in the subspace with
$m=0$, where $\gamma^2=1/4$. For $0<\alpha<1/2$ and
$1/2<\alpha<1$, repulsive potential appears in the subspaces with
$m=0$ and $m=-1$,  where $\gamma^2=1/4-\alpha^2$ and
$\gamma^2=1/4-(1-\alpha)^2$, respectively. Repulsive
P\"oschl-Teller system has no physical states with $E=0$. In
accordance with Eq. (\ref{ladder}), the continuous part of the
spectrum of the quantum system (\ref{Hquant}) with $\alpha\neq
\frac{1}{2}$ is described by the scattering states with $E>0$. The
corresponding eigenstates with $E=k^2$, $k>0$, can be expressed in
terms of the hypergeometric function. Any P\"oschl-Teller
subsystem (\ref{Htmal}) is characterized by a nonzero reflection
coefficient \cite{flugge}
\begin{equation}\label{Refl}
    |r|^2=\frac{1}{1+\rho^2},\quad \rho=\frac{\sinh \pi
    k}{\cos \pi \alpha}\, .
\end{equation}
In accordance with (\ref{Hh}), on the scattering states with $E>0$,
infinite-dimensional unitary irreducible representations of the
principal  continuous series of the algebra $sl(2,R)\sim so(2,1)$
 \cite{bargmann}
with $-\hat{J}_\mu\hat{J}^\mu=E+1/4>1/4$ and $j_3=\alpha+m$,
$m=0,\pm 1,\ldots$,  are realized.

System (\ref{Hquant}) with $\alpha\neq 1/2$  has also  bound
states of certain discrete negative energies. With the help of the
relation
\begin{equation}\label{HJ+J-J3}
    \hat{H}=\hat{J}_+\hat{J}_-
    -(\hat{J}_3-1/2)^2=\hat{J}_-\hat{J}_+ -(\hat{J}_3+1/2)^2,
\end{equation}
cf. (\ref{HmmD}) and (\ref{shape}), a part of corresponding
normalizable eigenstates is identified as the states annihilated
by the ladder operators $\hat{J}_-$ and $\hat{J}_+$. These are
\begin{equation}\label{j-E}
    \Psi^{\alpha,-}_{E,m}(\chi,\varphi)=e^{im\varphi}\cosh^{-(m+\alpha-1/2)}\chi,\quad
    \hat{J}_-\Psi^{\alpha,-}_{E,m}=0,\qquad
    E=-(m+\alpha-1/2)^2,
\end{equation}
where $m=1,2\ldots$ for $0\leq \alpha<1/2$, and $m=0,1,\ldots$ for
$1/2<\alpha<1$, and
\begin{equation}\label{j+E}
    \Psi^{\alpha,+}_{E,m}(\chi,\varphi)=e^{im\varphi}\cosh^{m+\alpha+1/2}\chi,\qquad
    \hat{J}_+\Psi^{\alpha,+}_{E,m}=0,\quad E=-(m+\alpha+1/2)^2,
\end{equation}
with $m=-1,-2,\ldots$ for $0\leq \alpha<1/2$, and $m=-2,-3,\ldots$
for $1/2<\alpha<1$. Infinite number of bound states with the same
energy eigenvalues are obtained then by the action of the ladder
operators $(\hat{J}_+)^n$ and $(\hat{J}_-)^n$, $n=1,2,\ldots$, on
the states (\ref{j-E}) and (\ref{j+E}). At $\alpha=0$ the discrete
part of the spectrum reveals a symmetry with respect to the change
$j_3\rightarrow -j_3$, but  it has no such a symmetry for
$\alpha\neq 0$. On the states
\begin{equation}\label{J+-n}
    (\hat{J}_+)^n\Psi^{\alpha,-}_{E,m}\quad  {\rm and}\quad
    (\hat{J}_-)^n\Psi^{\alpha,+}_{E,m},\quad n=0,1,\ldots,
\end{equation}
the half-bounded infinite-dimensional unitary representations of
the discrete series of the $sl(2,R)$ \cite{bargmann} are realized.
These representations are characterized by the value of the
Casimir operator and eigenvalues of the compact generator
$\hat{J}_3$, which are respectively
$-\hat{J}_\mu\hat{J}^\mu=-(m+\alpha)(m+\alpha-1)$, $j_3=m+\alpha
+n$, and $-\hat{J}_\mu\hat{J}^\mu=-(m+\alpha)(m+\alpha+1)$,
$j_3=m+\alpha -n$. The quantum spectrum of the system with
$\alpha=0$  is illustrated on Fig. 1. Note that the subspaces with
$j_3$ and $-j_3$, $j_3\neq 0$, are present symmetrically in the
spectrum, while the subspace with $j_3=0$ is unpaired.

\begin{figure}[h!]
\begin{center}
\includegraphics[width=0.6\linewidth]{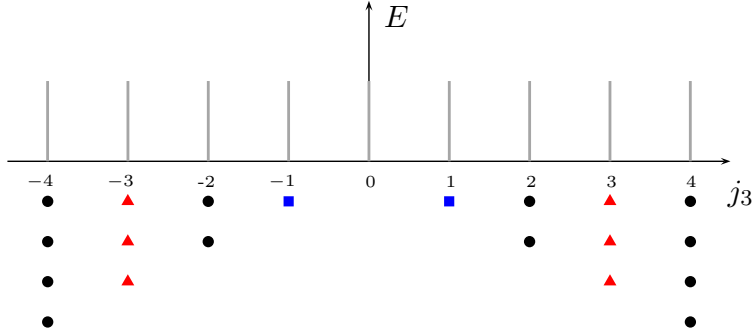}
\caption{Spectrum of the AdS${}_2$ system with integer Aharonov-Bohm
flux  $(\alpha=0)$.}\label{Fig1}
\end{center}
\end{figure}

Now, we will focus on the case $\alpha=1/2$. On the subspaces with
$m=0$ and $m=-1$, the dynamics reduces to that of the
one-dimensional free quantum particle, while the set of the
Hamiltonians (\ref{Htmal}) with all the possible integer values
$m_{1/2}=m$ corresponds to the family of the RPT Hamiltonians
(\ref{PT}) satisfying the identity (\ref{Hmm}),
\begin{equation}\label{Hj0m}
    \hat{H}\vert_{ j_3=\frac{1}{2}}=\hat{H}\vert_{
    j_3=-\frac{1}{2}}=H_0=-\partial_\chi^2,\qquad
    \hat{H}\vert_{ j_3=m+\frac{1}{2}}=H_m.
\end{equation}
The eigenstates $\Psi^{1/2}_{E,m}$ with $m=1,2\ldots$ and $E=k^2>0$
of the Hamiltonian (\ref{Hh}) are obtained from the free particle
plane wave states $e^{\pm ik\chi}$ ($m=0$) by the action on them of
the operator $(J_+)^m=e^{im\varphi}\mathcal{D}_{-m}
    {\cal D}_{-m+1}\ldots \mathcal{D}_{-1}$. This
corresponds exactly to relation (\ref{PTcont}). The scattering
states with negative values of $m$ are produced by the action of
the Hermitian conjugate operator $(J_-)^m$, $m=1,2\ldots$, on
$e^{\pm ik\chi}$. In comparison with the case $\alpha\neq 1/2$,
the system has additional states with $E=0$, on which the
half-bounded infinite-dimensional unitary representations of the
$sl(2,R)$ are realized. These states have a form of the
eigenvectors (\ref{J+-n}) constructed over the eigenstates of the
form (\ref{j-E}) and (\ref{j+E}) with $\alpha=1/2$, in which $m=0$
and $m=-1$, respectively. The normalizable negative energy states
are constructed in the same way over the eigenstates (\ref{j-E})
and (\ref{j+E}) with $m=1,2\ldots$ and $m=-2,-3,\ldots$. In this
case all the energy levels in the spectrum shown on Fig. 2 have a
double degeneration  with respect to the reflection
$j_3\rightarrow -j_3$, cf. Fig. 1.

\begin{figure}[h!]
\begin{center}
\includegraphics[width=0.6\linewidth]{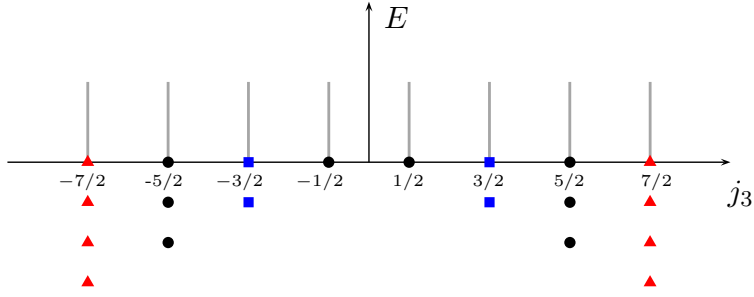}
\caption{Spectrum of the AdS${}_2$ system with half-integer
Aharonov-Bohm flux  ($\alpha=1/2$).}\label{figS}
\end{center}
\end{figure}

In terms of the ladder operators, second involutive automorphism
from (\ref{autoSO}) takes the form
\begin{equation}\label{SJJJ}
    S:\, (\hat{J}_{\pm}, \hat{J}_{3}) \rightarrow (-\hat{J}_{\mp},-\hat{J}_3).
\end{equation}
Under this transformation quantum symmetry algebra (\ref{so21}) and
Hamiltonian (\ref{Hh}) are invariant.
 The described properties of the states of the quantum system mean that this
discrete symmetry is spontaneously broken except the cases when the
Aharonov-Bohm flux takes integer and half-integer values. In the
case $\alpha=0$ or $\alpha=1/2$, this discrete symmetry transforms
mutually the half-bounded infinite-dimensional representations of
the $so(2,1)$ realized on the states of the discrete spectrum
(including the states $E=0$ when $\alpha=1/2$).

\section{Reduction and tri-supersymmetry}
Let us  identify now the exact analogs of the Crum-Darboux
intertwining relations, and trace out how the tri-supersymmetric
structure appears under the appropriate reduction of the AdS${}_2$
system in the presence of the Aharonov-Bohm effect with the
half-integer flux ($\alpha=1/2)$.  In conclusion of this section we
shall comment on the case of the integer flux.

Having in mind factorization (\ref{psiEm}), we introduce notations
$|m\rangle=e^{im\varphi}$ and $\langle
\psi_1|\psi_2\rangle=\frac{1}{2\pi}\int_0^{2\pi}\psi_1^*\psi_2
d\varphi$, so that $\langle m|m'\rangle=\delta_{mm'}$. Taking into
account explicit form of the ladder operators (\ref{J+-}), we get
the only nonzero  matrix elements of $\hat{J}_\pm$ and $\hat{H}$,
\begin{eqnarray}\label{DDJJ}
    &\langle m-1|\hat{J}_-|m\rangle=-{\cal D}_m,\quad
    \langle m|\hat{J}_+|m-1\rangle={\cal D}_{-m},&\\
    &\langle m|\hat{H}|m\rangle=-{\cal D}_{-m}{\cal D}_m-m^2,&\label{HDD}
\end{eqnarray}
that correspond to (\ref{D}) and (\ref{HmmD}). Then, with taking
into account (\ref{DDJJ}) and (\ref{HDD}), the Crum-Darboux
intertwining relation  (\ref{CD1}) is just the matrix element of the
relation $[(\hat{J}_-)^{l+1},\hat{H}]=0$, $l=0,1,\ldots$,
\begin{equation}\label{CDmatr}
    \langle m-l-1|\hat{J}_-^{l+1}\hat{H}|m\rangle=\langle
    m-l-1|\hat{H}\hat{J}_-^{l+1}|m\rangle.
\end{equation}
The intertwining relation (\ref{CD}) is given  by Hermitian
conjugation, or, alternatively, is produced by the same operator
relation taken between the states $\langle -m-1|$ and
$|-m+l\rangle$ with subsequent use of the identity (\ref{Hmm}).
Relation (\ref{CDmatr}) taken for $l=2m$ (or, its Hermitian
conjugate) together with the identity (\ref{Hmm}) produce Eq.
(\ref{calA}) corresponding to the nontrivial odd order integral of
the RPT system $H_m$.

Hence, we have identified the origin of the intertwining relations
associated with the RPT Hamiltonian (\ref{PT}). Since the
commutativity of the operators ${\cal X}_{m,l}$ and ${\cal
Y}_{m,l}$ with Hamiltonian ${\cal H}_{m,l}$ is reduced to the
intertwining relations, one can find what matrix elements of the
operators $\hat{J}_\pm^r$, with appropriately chosen integer
$r>0$, correspond to the operators $X^\pm_{m,l}$, $Y^\pm_{m,l}$
and $Z^\pm_{m,l}$, from which the integrals ${\cal X}_{m,l}$,
${\cal Y}_{m,l}$ and ${\cal Z}_{m,l}$ are composed. In
correspondence with the discrete symmetry (\ref{SJJJ}) we have
\begin{eqnarray}\label{X-matrix}
    -(-i)^{m-l}\langle l|   \hat{J}_-^{m-l}|m\rangle=
    -i^{m-l}\langle -l-1|\hat{J}_+^{m-l}|-m-1\rangle&=&X^-_{m,l}\, ,\\
\label{Y-matrix}
    -(-i)^{m+l+1}\langle -l-1|   \hat{J}_-^{m+l+1}|m\rangle=
    -i^{m+l+1}\langle l|\hat{J}_+^{m+l+1}|-m-1\rangle&=&Y^-_{m,l}\, ,\\
\label{Z-matrix}
    (-i)^{2m+1}\langle -m-1|   \hat{J}_-^{2m+1}|m\rangle=
    i^{2m+1}\langle m|\hat{J}_+^{2m+1}|-m-1\rangle&=&Z^-_{m,l}\, ,\\
\label{Z+matrix}
    (-i)^{2l+1}\langle
    -l-1|\hat{J}_+^{m-l}\hat{J}_-^{m+l+1}|l\rangle=
     i^{2l+1}\langle l|\hat{J}_+^{m+l+1}
     \hat{J}_-^{m-l}|-l-1\rangle&=&
   Z^+_{m,l}\, .
\end{eqnarray}
Operators $X^+_{m,l}$ and $Y^+_{m,l}$ are obtained by Hermitian
conjugation of (\ref{X-matrix}) and (\ref{Y-matrix}). These
relations are illustrated by Figures 3--7 for $m=3$, $l=1$.

\begin{figure}[h!]
\begin{center}
\includegraphics[width=0.6\linewidth]{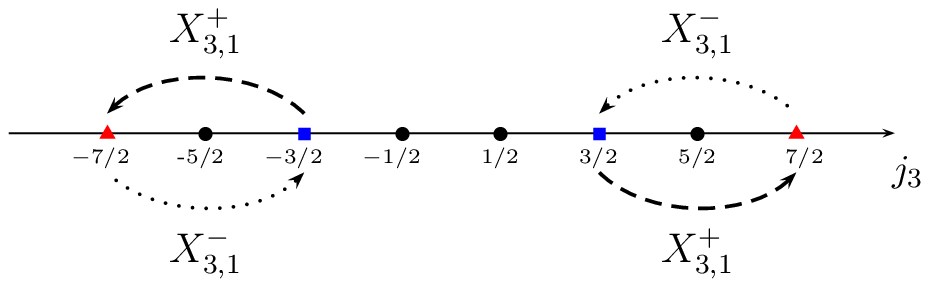}
\caption{Action of the operators $X^\pm_{3,1}$ in correspondence
with (\ref{X-matrix}).}\label{figX}
\end{center}
\end{figure}

\begin{figure}[h!]
\begin{center}
\includegraphics[width=0.6\linewidth]{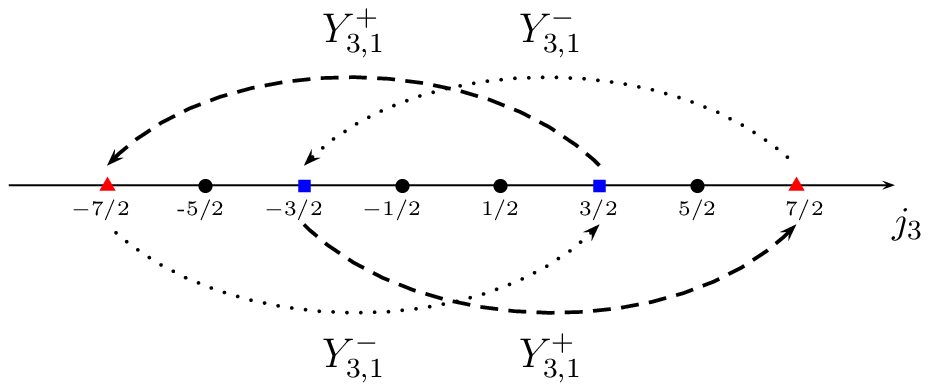}
\caption{Action of the operators $Y^\pm_{3,1}$ in correspondence
with (\ref{Y-matrix}).}\label{figY}
\end{center}
\end{figure}

\begin{figure}[h!]
\begin{center}
\includegraphics[width=0.6\linewidth]{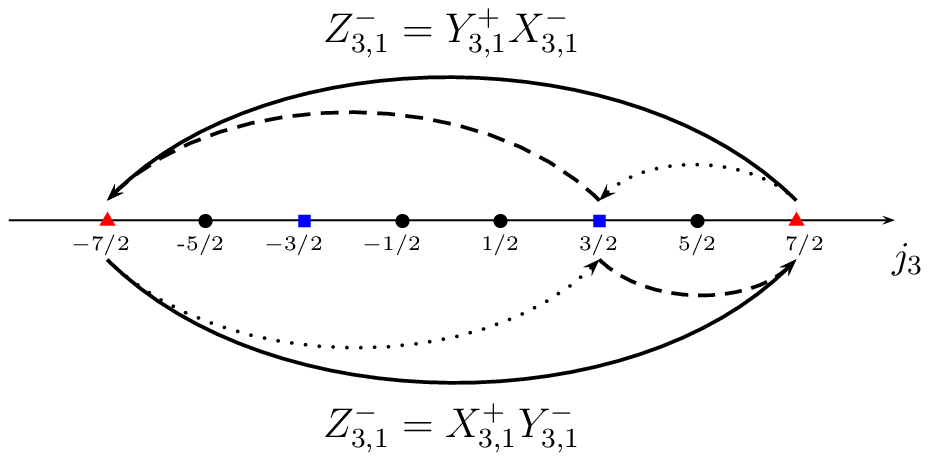}
\caption{Action of the operator $Z^-_{3,1}$ in correspondence with
(\ref{Z-matrix}).}\label{figZ}
\end{center}
\end{figure}

\begin{figure}[h!]
\begin{center}
\includegraphics[width=0.6\linewidth]{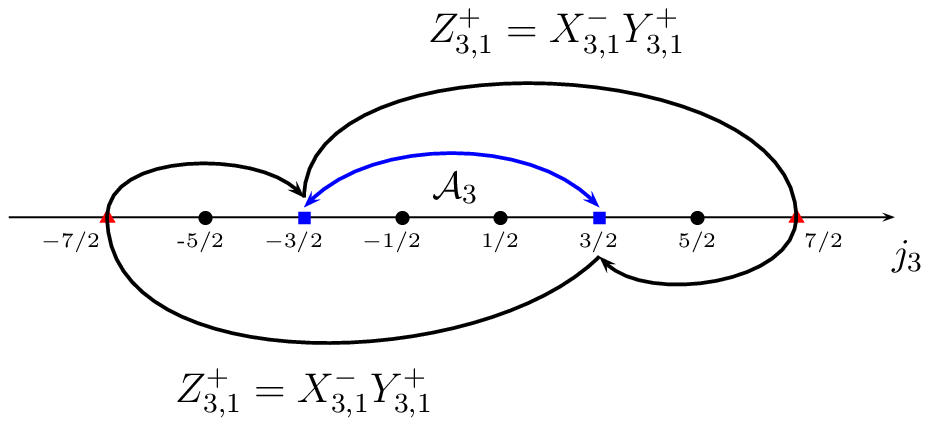}
\caption{Action of the operator $Z^+_{3,1}$ in correspondence with
(\ref{Z+matrix}).}\label{figZp}
\end{center}
\end{figure}

In order to reproduce the relations corresponding to (\ref{XXYY})
with the help of (\ref{X-matrix})--(\ref{Z+matrix}), we can use
the identities
 \begin{equation}
    \hat{J}_{+}^n\hat{J}_{-}^n=\prod_{k=0}^{n-1}
    \left(\hat{H}+\left(\hat{J}_3-k-\frac{1}{2}\right)^2
    \right)\equiv P_n(\hat{H},\hat{J}_3),\qquad
    \hat{J}_{-}^n\hat{J}_{+}^n=
    P_n(\hat{H},-\hat{J}_3),\label{identity}
 \end{equation}
that follow from the $so(2,1)$ algebra and Eq. (\ref{HJ+J-J3}).
For instance, putting in the first identity $n=2m+1$, and
computing a diagonal matrix element between the states $\langle
m|$ and $|m\rangle$, we reproduce the  lower component case of the
third relation from (\ref{XXYY}), i.e.
$(Z^-_{m,l})^2=P_{2m+1}(H_m)$, where the polynomial is given by
Eq. (\ref{P2m+a}).

Therefore,  the reduction of the AdS${}_2$ system with
half-integer Aharonov-Bohm flux ($\alpha=1/2$) to an eigensubspace $j_3=m+\frac{1}{2}$ reproduces a bosonized
nonlinear supersymmetry of the RPT system. On the Hilbert space
composed from the two eigenspaces $j=m+\frac{1}{2}$ and
$j=l+\frac{1}{2}$ with $m\neq l$ we reveal the tri-supersymmetric
structure of the extended RPT system (\ref{superPT}). The key role
for the nontrivial nonlinear supersymmetric structure is played
here by the involutive automorphism (\ref{SJJJ}) of the $so(2,1)$
algebra, which is realized as a symmetry on the Hilbert space of
the two-dimensional quantum system.

In the previous section we have seen that this discrete symmetry is
not spontaneously broken also for integer values of the magnetic
flux. In this case, however, the system is not reflectionless, and
cannot be related to the one-dimensional quantum free particle
system. Nevertheless, the intertwining relations corresponding to
the Crum-Darboux transformations can  be considered in this case as
well, and they also can be related with the ladder operators. Using
the intertwining relations for the case of half-integer values of
the index $m=n+\frac{1}{2}$, $n\in \Z$, and the symmetry
(\ref{Hmm}), we find that the operator ${\cal D}_{-m}{\cal
D}_{-m+1}\ldots {\cal D}_{-1/2}{\cal D}_{1/2}\ldots {\cal
D}_{m-1}{\cal D}_m$ of \emph{even} order $2m+1=2(n+1)$ commutes with
the Hamiltonian $H_m$. A simple calculation shows, however, that
this operator reduces to a certain polynomial  of order $n+1$ in
$H_{n+\frac{1}{2}}$. For instance, in correspondence with
(\ref{HmmD}), ${\cal D}_{-1/2}{\cal D}_{1/2}=-H_{1/2}-1/4$. Hence,
there is no bosonized and tri- supersymmetric structure in the
P\"oschl-Teller system with a half-integer value of the parameter
$m$.


\section{On (super)conformal dynamical symmetry}

As we observed in section 2, the nonlinear supersymmetry of the RPT
system is generated by the Crum-Darboux transformations from the
hidden bosonized $N=2$ linear supersymmetry of the free
non-relativistic particle. The latter system possesses also a
\emph{non-relativistic conformal symmetry} in the form of a
dynamical $so(2,1)$ symmetry \cite{NiedHag}. The same dynamical 
symmetry is present in some other quantum mechanical
non-relativistic systems, including, in particular, the conformal
mechanics model \cite{ConfMod}, and the model of a particle in the
field of a magnetic vortex \cite{Jac2,susyvort}. The latter
corresponds to the planar Aharonov-Bohm effect. The conformal
mechanics model, like the RPT system, is related to the free
particle by a Crum-Darboux transformation \cite{COP}.  The conformal
$so(2,1)$ dynamical symmetry of all these systems can be explained
within the framework of the AdS/CFT correspondence, see \cite{LP1}
and \cite{DHH}.

In this section, we construct analogs of the free particle
generators of the $so(2,1)$ symmetry for the RPT system, and
investigate their properties. The hidden bosonized $N=2$ linear
 supersymmetry of the free particle can be unified with its dynamical
$so(2,1)$ symmetry to produce a linear bosonized superconformal
$osp(2|2)$ symmetry, see \cite{COP}. We also discuss a realization
of the $osp(2|2)$ supersymmetry in the RPT system.

Before we pass over to the discussion of the specified problem, let
us make two notes. The non-relativistic AdS/CFT correspondence
appears between a relativistic theory in the AdS space-time of $d+3$
dimensions, and non-relativistic theories in $(d+1)$-dimensional
space-time \cite{DHH}. We exploited  here the relation between  the
non-relativistic particle system on the AdS${}_2$, whose evolution
is described by an external time variable, and the non-relativistic
$(1+1)$-dimensional RPT system. Next, in the AdS${}_2$ system, the
$so(2,1)$ isometry corrected by the magnetic flux is represented by
the true, time-independent integrals of motion. Hamiltonian
(\ref{HJ+J-J3}) is the Casimir operator of this $so(2,1)$ symmetry,
whose generators are differential operators of the first order. In
contrary, the generators of the conformal dynamical symmetry of the
free particle we consider below are represented by differential
operators of the second order.

The free particle is characterized by the Schr\"odinger symmetry
\cite{NiedHag} Lie algebra with the following nonzero commutation
relations,
\begin{equation}\label{5-3}
    [D_0,H_0]=2iH_0, \quad [H_0,K_0]=-4iD_0, \quad [D_0,K_0]=-2i K_0.
\end{equation}
\begin{equation}\label{DPB}
    [B_0,P_0]=i\cdot 1,\quad
    [D_0,P_0]=i P_0,\quad
    [D_0,B_0]=-i B_0,
\end{equation}
\begin{equation}\label{PBHK}
    [B_0,H_0]=2iP_0,\quad
    [P_0,K_0]=-2iB_0,
\end{equation}
see also \cite{superS}. Here
\begin{equation}\label{defPBDK}
    P_0=-i\frac{d}{dx},\quad
    B_0=x -2P_0t,\quad
    H_0=P_0^2,\quad
    D_0=\frac{1}{2}\{P_0,B_0\},\quad
    K_0=B_0^2.
\end{equation}
In  the first relation from (\ref{DPB}) the unit operator
corresponds to the mass put here equal to one-half, which is a
central element of the algebra. The boost generator $B_0$, and the
generators of dilatations, $D_0$, and special conformal
transformations (expansions), $K_0$, depend explicitly on time.
This means that the Schr\"odinger symmetry is \emph{dynamical}.
Its subalgebra (\ref{5-3}) is the conformal $so(2,1)$ dynamical
symmetry.

Reflection operator $R$, being integral of motion, anticommutes
not only with the $P_0$, but also with the time-dependent integral
$B_0$. Therefore, the free particle can  be characterized by the
dynamical $osp(2|2)$ superconformal symmetry as well. Its
superalgebra is given by the $so(2,1)$ commutation relations
(\ref{5-3}) supplied with the nontrivial (anti)-commutation
relations
\begin{equation}\label{ospfree}
    \{Q^{(0)}_a,Q_b^{(0)}\}=2\delta_{ab}H_0,\quad
    \{S^{(0)}_a,S_b^{(0)}\}=2\delta_{ab}K_0,\quad
    \{Q^{(0)}_a,S_b^{(0)}\}=2\delta_{ab}D_0-\epsilon_{ab}\Sigma,
\end{equation}
\begin{equation}\label{HSQD}
    [H_0,S^{(0)}_a]=-2iQ_a^{(0)},\quad
    [K_0,Q^{(0)}_a]=2iS_a^{(0)},
\end{equation}
\begin{equation}\label{DQS0}
    [D_0,Q^{(0)}_a]=iQ_a^{(0)},\quad
     [D_0,S^{(0)}_a]=-iS_a^{(0)},
\end{equation}
\begin{equation}\label{SigmaQS}
    [\Sigma,Q^{(0)}_a]=2i\epsilon_{ab}Q^{(0)}_b,\quad
    [\Sigma,S^{(0)}_a]=2i\epsilon_{ab}S^{(0)}_b.
\end{equation}
Here
\begin{equation}\label{QSDSnot}
    Q_1^{(0)}=P_0,\quad
    Q_2^{(0)}=iRQ_1^{(0)},\quad
    S_1^{(0)}=B_0,\quad
    S_2^{(0)}=iRS_1^{(0)},\quad
    \Sigma=-R.
\end{equation}

Suppose now that we have two quantum almost isospectral systems
related by the Crum-Darboux intertwining relations (\ref{tilH}). Let
a system $H$ possesses an integral of motion ${\cal A}$ which can be
time dependent, $\frac{d}{dt}{\cal A}=\frac{\partial}{\partial
t}{\cal A}-i[{\cal A},H]=0$. Time-independence of the Crum-Darboux
generator $A_n$, and the intertwining relations (\ref{tilH}) allow
us to construct an analog of the integral ${\cal A}$ for the system
$\tilde{H}$,
\begin{equation}\label{calAtil}
    \tilde{\cal A}=A_n{\cal A}A_n^\dagger,\qquad \frac{d}{dt}\tilde{\cal A}
    =\frac{\partial}{\partial t}\tilde{\cal A}-i[\tilde{\cal A},\tilde{H}]=0.
\end{equation}
It depends on time if and only if ${\cal A}$ is time dependent as
well. Identifying $H$ and $\tilde{H}$ with $H_0$ and $H_m$, and
$A_n$ with the operator ${\cal D}_{-m}\ldots {\cal D}_{-1}\equiv
X_m^\dagger =-i^mX^\dagger_{m,0}$, see Eqs. (\ref{D}) and
(\ref{DX}), we find the analogs of the even, $D_0$, $K_0$, and odd,
$Q^{(0)}_a$, $S^{(0)}_a$, free particle integrals for the RPT system
described by the Hamiltonian (\ref{PT}). We denote these analogs
just by changing the index $0$ for $m$.  The integral $\Sigma$,
defined in (\ref{QSDSnot}), is also the integral for the system
$H_m$.

Let us look now what happens with the dynamical $so(2,1)$ symmetry
in the RPT system. A direct computation gives the commutation
relations
 \begin{equation}\label{HDHm}
     [D_m,H_m]=2iH_mP_m(H_m),\quad [K_m,H_m]=4iD_m
 \end{equation}
where in correspondence with (\ref{XXYY}), (\ref{PXH}),
$P_m(H_m)=X^{\dagger}_mX_m=\prod_{n=0}^{m-1}(H_m-E_{m;n})$. To find
$[D_m,K_m]$, we use the identity $D_0^2=H_0K_0+2iD_0+\frac{3}{4}$,
and get
\begin{equation}\label{5-7}
 [D_m,K_m]=a_m(H_m)K_m+b_m(H_m)D_m+c_m(H_m)
\end{equation}
where $a_m$, $b_m$ and $c_m$ are some polynomials in $H_m$. At this
level,  like in the case of the nonlinear supersymmetry analyzed in
the previous sections, the Lie algebra of the conformal symmetry of
the free particle (\ref{5-3}) is just deformed by the polynomials in
the Hamiltonian of the corresponding RPT system. The complete
algebraic structure, however, is more complicated here. Unlike the
supercharges $Q^{(m)}_a=Z_a$, the dynamical integrals $D_m$ and
$K_m$ do not commute with the Hamiltonian $H_m$. As a consequence,
the repeated commutators produce polynomials in all the three
generators $H_m$, $D_m$, $K_m$.  To illustrate this phenomenon,
consider the simplest case of $m=1$.  The reflectionless
P\"oschl-Teller Hamiltonian $H_1=-{d^2}/{dx^2}-{2}/{\cosh^2 x}$ is
intertwined with the free particle Hamiltonian by the first order
differential operator $X_1=\mathcal{D}_1=d/dx+\tanh x$. The
operators $D_1$, $H_1$ and $K_1$ satisfy the commutation relations

\begin{equation}\label{m1DHK}
[K_1,H_1]=4iD_1\, ,
\end{equation}
\begin{equation}
[D_1,H_1]=2iH_1(H_1+1) ,\qquad [D_1,K_1]=-2iK_1-2i
\{K_1,H_1\}-3i(H_1+1)\,.
\end{equation}
Denoting $\mathcal{K}_1\equiv[D_1,K_1]$, we find
\begin{equation}
    [\mathcal{K}_1,K_1]\equiv\mathcal{M}_1=-8\{K_1,D_1\} -12D_1,
\end{equation}
and
\begin{equation}\label{m1DHK2}
     [\mathcal{M}_1,K_1]=32i\{K_1^2,H_1\}+32iK_1^2+176i\{K_1,H_1\}+200iK_1+228i(H_1+1).
\end{equation}
{}From the displayed  structure it is clear that despite the
nonlinear character of the algebra generated by the $H_1$, $D_1$ and
$K_1$, all the (repeated) commutators are polynomials in them. No
new independent integrals do appear.

This ceases to be true once we incorporate the fermionic analogs of
the $osp(2|2)$ generators.  Besides the deformation effect we
observed above, the commutation relations between bosonic and
fermionic operators produce new fermionic dynamical integrals, not
reducible to the products of the fermionic generators $Q^{(m)}_a$,
$S^{(m)}_a$, and of the bosonic generators. Nevertheless, the number
of new fermionic integrals is finite. It depends  on the value of
the parameter $m$, and therefore, on the order of the Crum-Darboux
generating operator $X_m$. To illustrate this phenomenon, again,
restrict ourselves by the simplest case of the RPT system with
$m=1$. In addition to the already specified (repeated) commutation
relations between bosonic operators $H_1$, $D_1$ and $K_1$, we get
the following  (anti)-commutation relations,

\begin{equation}\label{Sosp1}
    \{Q^{(1)}_a,Q^{(1)}_b\}=2\delta_{ab}H_1(H_1+1)^2,\quad
    \{S^{(1)}_a,S^{(1)}_b\}=\delta_{ab}
    \left(\{K_1,H_1\}+2(K_1+H_1+1)    \right),
\end{equation}
\begin{equation}
    \{S^{(1)}_a,Q^{(1)}_b\}=\delta_{ab}\left(\{D_1,H_1\}+
    2D_1\right)-\epsilon_{ab}\Sigma(3H_1+1)(H_1+1),
    \label{Sosp2}
\end{equation}
\begin{equation}
    \{J^{(1)}_a,J^{(1)}_b\}=\frac{1}{8}\delta_{ab}
    \left(K_1H_1K_1-\{K_1,H_1\}-2(H_1+1)
    \right),
    \label{Sosp3}
\end{equation}
\begin{equation}
     \{S^{(1)}_a,J^{(1)}_b\}=\frac{1}{4}\delta_{ab}
     \left(\{K_1,D_1\}+2D_1
     \right)+\frac{1}{8}\epsilon_{ab}\Sigma( \{K_1,H_1\}-2K_1+2(H_1+1)),
    \label{Sosp4}
\end{equation}
\begin{equation}
     \{J^{(1)}_a,Q^{(1)}_b\}=\frac{1}{4}\delta_{ab}
     \left(2H_1K_1H_1+
     \{K_1,H_1\}-5H_1^2-4H_1+1
     \right)
     -\frac{1}{4}\epsilon_{ab}
     \Sigma\left( 3\{D_1,H_1\}+2D_1  \right),
    \label{Sosp5}
\end{equation}
\begin{equation}
    [K_1,Q^{(1)}_a]=3i\{S^{(1)}_a,H_1\}+2iS^{(1)}_a,
    \quad[K_1,S^{(1)}_a]=8iJ^{(1)}_a,
    \quad[K_1,J^{(1)}_a]=\frac{i}{4}\{K_1,S^{(1)}_a\},
    \label{Sosp6}
\end{equation}
\begin{equation}
    [H_1,Q^{(1)}_a]=0, \quad [H_1,S^{(1)}_a]=-2iQ^{(1)}_a,
    \quad [H_1,J^{(1)}_a]=-\frac{i}{2}\{S^{(1)}_a,H_1\},
    \label{Sosp7}
\end{equation}
\begin{equation}
    [D_1,Q^{(1)}_a]=iQ^{(1)}_a\left(1+3H_1
    \right),\quad
    [D_1,S^{(1)}_a]=-\frac{i}{2}\left(2S^{(1)}_a +\{S^{(1)}_a,H_1\} \right)
    \label{Sosp8}
\end{equation}
\begin{equation}
    [D_1,J^{(1)}_a]=-\frac{3i}{2}\{J^{(1)}_a,H_1\}-iJ^{(1)}_a-iQ^{(1)}_a,
    \label{Sosp9}
\end{equation}
\begin{equation}
    [\Sigma,Q^{(1)}_a]=2i\epsilon_{ab}Q^{(1)}_b,\quad
    [\Sigma,S^{(1)}_a]=2i\epsilon_{ab}S^{(1)}_b, \quad
    [\Sigma,J^{(1)}_a]=2i\epsilon_{ab}J^{(1)}_b.
    \label{Sosp10}
\end{equation}
Here
\begin{equation}\label{Jnew}
    J^{(1)}_1=X^\dagger_1 J^{(0)}_1 X_1,\quad
    J^{(1)}_2=iRJ^{(1)}_1,
\end{equation}
are the two new fermionic  integrals generated by the commutator of
$K_1$ with $S^{(1)}_a$, see (\ref{Sosp6}). They correspond to the
dynamical integral $J^{(0)}_1=\frac{1}{8}\{K_0,Q^{(0)}_1\}$ of the
free particle. {}From the point of view of the closed Lie
superalgebraic structure of the free particle model, the operator
$J^{(0)}_1$ belongs to the universal enveloping algebra of the
$osp(2|2)$.

The displayed set of the anti-commutation relations shows,
nevertheless,  that no other new independent integrals will be
generated by the repeated anticommutation relations. Note that a
similar picture corresponding to the deformation and extension of
the superconformal $osp(2|2)$ symmetry appears in the
fermion-monopole system \cite{fermimono}. This happens also in the
superconformal mechanics model when a classical boson-fermion
coupling constant $\alpha$ is changed for $n\alpha$, where  $n>1$ is
integer, see \cite{LPA}.

We do not touch here the question of relation of the deformed (and
extended) conformal $so(2,1)$ (and superconformal $osp(2|2)$)
dynamical symmetry of the RPT system to the $so(2,1)$ symmetry of
the particle on the AdS${}_2$. Let us note only that the generators
of the conformal and superconformal dynamical symmetries of the RPT
system could be related to  the oscillator-like operators
$P_+=e^{i\varphi}\hat{J}_+$ and $P_-=P_+^{\dagger}
=\hat{J}_-e^{-i\varphi}$ of the AdS${}_2$ system. These two
operators commute with the operator $\hat{J}_3$, and represent the
observable operators under reduction to the one-dimensional RPT
system. The operators $P_\pm$  are not integrals of motion for the
original two-dimensional system due to a nontrivial dynamics of the
angular variable $\varphi$. Nevertheless, the true and also
dynamical integrals of motion of the RPT system can be obtained from
them after reduction. Investigation of this aspect lies, however,
beyond the scope of the present paper.

\section{Discussion and outlook}

We showed that the hidden bosonized nonlinear supersymmetry of the
reflectionless P\"oschl-Teller system and tri-supersymmetric
structure of the pair of the RPT systems have an origin in the
Aharonov-Bohm effect for non-relativistic particle on the AdS${}_2$.
Both supersymmetric structures are based on the two involutive
automorphsims of the $so(2,1)$ algebra of the AdS${}_2$ isometry
with generators corrected by the magnetic flux $\alpha$. One of
these automorphisms, $R$, corresponds to a reflection of a
space-like coordinate of the ambient Minkowski space. It is the
symmetry for any value of the flux. Another automorphism, $S$,
corresponds to a reflection of one of the time-like coordinates of
the ambient space. This discrete symmetry is unbroken only in the
case of integer and half-integer values of the flux.

Classically,  the system is subject to the P\"oschl-Teller
potential $V(\chi;J_3)=-J_3^2\cosh^{-2}\chi$ in the noncompact
AdS${}_2$ coordinate $\chi$. It acquires a correction term in the
quantum framework, $V=-(\hat{J}_3^2-\frac{1}{4})\cosh^{-2}\chi$.
The classical free-particle dynamics in $\chi$ takes place  for
zero value of the axial angular momentum shifted by the magnetic
flux, $J_3=0$. It can be recovered on the quantum level for
half-integer values of the magnetic flux only. In this case, the
dynamics of any partial wave of the particle on the AdS${}_2$ is
governed by the reflectionless P\"oschl-Teller potential, which
turns into zero in two subspaces $j_3=\pm \frac{1}{2}$. The
complete two-dimensional quantum dynamics on the AdS${}_2$ is
hence reflectionless. This is not the case for the other values of
the flux.  For $\alpha\neq \frac{1}{2}+n$, the dynamics of some
partial waves is subject to the repulsive potential.  Then this
information is transmitted to all other partial wave sectors by
the $so(2,1)$ ladder operators, which play the role of the
generators of the Crum-Darboux transformations. So, the
two-dimensional dynamics ceases to be reflectionless in this case.

For half-integer values of the flux, all the spectrum of  the
two-dimensional quantum system acquires an additional double
degeneration related to the Hamiltonian symmetry
$\hat{J}_3\rightarrow -\hat{J}_3$  generated by the $S$. Such a
degeneration is absent for the unpaired partial wave sector $j_3=0$
in the case of integer flux. It is the additional double
degeneration that is behind the existence of the nontrivial
nonlinear supersymmetric structure in the RPT system appearing after
reduction by the compact $so(2,1)$ generator $\hat{J}_3$. Generator
of the other discrete symmetry, $R$, is transformed then into the
$\Z_2$-grading operator of the hidden bosonized supersymmetry of the
RPT system. The supersymmetry generators of the reduced system
correspond to certain powers of the $so(2,1)$ ladder operators of
the original two-dimensional system with half-integer flux.

Note that the special ``magic" of half-integer values of the flux
and related degeneracy were discussed in a different context in
\cite{AhCol}.

The described correspondence between the non-relativistic AdS${}_2$
Aharonov-Bohm effect and the one-dimensional P\"oschl-Teller system
for a generic magnetic flux value case is somewhat of the AdS/CFT
holography \cite{ads1} nature. This observation deserves a deeper
study that can be useful in the context of the integrable nonlinear
equations.

For the peculiar case of half-integer values of the flux, we
considered this aspect in the context of non-relativistic AdS/CFT
\cite{NRCFT,DHH}. Namely, we showed, that the analogs of the free
particle generators of the dynamical conformal, $so(2,1)$, and
bosonized superconformal, $osp(2|2)$, symmetries can be transferred
to the RPT system  by the Crum-Darboux transformation. In the case
of the conformal $so(2,1)$ dynamical symmetry, there appears a
certain nonlinear deformation of the $so(2,1)$ algebra. In the case
of the superconformal  $osp(2|2)$ symmetry, in addition to a
deformation, finite number of new fermionic integrals is generated.
These dynamical symmetries of the RPT system need a further
investigation.

The conformal $so(2,1)$ dynamical symmetry of the planar
Aharanov-Bohm effect finds a natural explanation in the AdS/CFT
context  \cite{LP1,DHH}. The bound state Aharonov-Bohm effect, that
corresponds to a particle confined to a circle pierced by the
Aharonov-Bohm flux, is characterized by the hidden bosonized
supersymmetry as well \cite{PTsusy}. It would be interesting to look
for the hidden bosonized $N=2$ supersymmetry and bosonized
superconformal $osp(2|2)$ dynamical symmetry in the planar
Aharonov-Bohm effect \cite{CFJP}.

Nonlinear bosonized  supersymmetry of the RPT system and
tri-supersymmetry of the extended RPT system are related to the
nontrivial Lax operator which is the higher order differential
operator. As we saw, this operator is the first order integral
$\frac{d}{dx}$ of the free particle  transferred to the RPT system
by the Crum-Darboux transformation. Its square  gives a spectral
polynomial of the RPT system, that has double roots and describes a
degenerate hyperelliptic curve \cite{integrable}. This degeneration
originates from the infinite-period limit applied to the finite-gap
periodic Lam\'e, or associated Lam\'e system. It is this limit that
produces the RPT system with its unique, pure imaginary period from
the double periodic model. The spectral polynomial of the associated
Lam\'e system is non-degenerate. It was showed in
\cite{trisusy,trisusy1} that such a finite-gap periodic system is
also characterized by the hidden bosonized nonlinear supersymmetry,
while the pair of such systems is described by the
tri-supersymmetric structure. In the case of the finite-gap periodic
systems, Crum-Darboux transformations relate the systems with the
same number of gaps. Hence, an $n$-gap  ($n\in\N$) periodic system
can not be related by Crum-Darboux transformation to the quantum
free particle model, which corresponds to the simplest case of a
zero-gap periodic system. Then a question appears: is there any
analog of the Aharonov-Bohm effect explanation for a hidden
bosonized supersymmetry and related tri-supersymmetric structure in
the finite-gap periodic systems? We are going to investigate this
question elsewhere.

The authors thank Peter Horvathy for stimulating communications.
The work has been partially supported by CONICYT, DICYT (USACH)
and by FONDECYT under grants 1050001 and 3085013.

\end{document}